\documentclass[aps,prd,superscriptaddress,showpacs,secnumarabic, eqsecnum]{revtex4}
\usepackage{graphicx}
\usepackage{dcolumn}
\usepackage{bm}

\newcommand{\ba}{\begin{eqnarray}}
\newcommand{\ea}{\end{eqnarray}}
\def\be{\begin{equation}}
\def\ee{\end{equation}}
\def\n{\nonumber\\}
\def\x{{\rm x}}

\begin{document}

\title{Incidence of the boundary shape in the effective theory of  fractional quantum Hall edges}

\author{D.C.\ Cabra}
\email{cabra@lpt1.u-strasbg.fr}
\affiliation{
Laboratoire de Physique Th\'eorique, Universit\'e Louis
Pasteur, 3 rue de l'Universit\'e, F-67084 Strasbourg Cedex, France
~~\\
Departamento de F\'{\i}sica, Universidad Nacional de la Plata,
C.C.\ 67, (1900) La Plata, Argentina
~~\\
Facultad de Ingenier\'\i a,  Universidad Nacional de Lomas de
Zamora, Cno. de Cintura y Juan XXIII, (1832) Lomas de Zamora,
Argentina.
~~\\}

\author{N.\ E.\ Grandi}
\email{grandi@fisica.unlp.edu.ar}
\affiliation{Departamento de F\'{\i}sica, Universidad Nacional de la Plata,
C.C.\ 67, (1900) La Plata, Argentina}

\date{\today}

\begin{abstract}
Starting from a microscopic description of a system of strongly
interacting electrons in a strong magnetic field in a finite
geometry, we construct the boundary low energy effective theory for
a fractional quantum Hall droplet taking into account the effects of
a smooth edge. The effective theory obtained is the standard chiral
boson theory (chiral Luttinger theory) with an additional
self-interacting term which is induced by the boundary. As an
example of the consequences of this model, we show that such
modification leads to a non-universal reduction in the tunnelling
exponent which is independent of the filling fraction. This is in
qualitative agreement with experiments, that systematically found
exponents smaller than those predicted by the ordinary chiral
Luttinger liquid theory.
\end{abstract}


\maketitle

\section{Introduction}

Quantum Hall effect (QHE) has been the subject of intense research
since 1980 when the first plateaux in the Hall conductivity were
observed for integer filling fractions \cite{von Klitzing}. The
interest on these systems was then reinforced after the fractional
QHE (FQHE) was identified as a novel state of matter \cite{FQHE}.
Since then, different approaches have been developed and applied to
the study of these systems, such as the successful variational
ground-state wave function proposed by Laughlin \cite{Laughlin}.
This description led to the concept of fractionally charged
quasiparticles as a consequence of the strong correlations in the
system, which were in turn experimentally observed \cite{fractional
exp}. The composite fermion approach \cite{Jain}, which provides a
description of the FQHE as a QHE where a quasiparticle is built up
as a bound state of an electron and a given number of magnetic flux
quanta, also provides a consistent description. An alternative
description of these systems, based on the Chern-Simons topological
action also reproduces many of the experimentally observed features
\cite{Girvin}.

In spite of the enormous success of all these descriptions, a theory
for the QHE derived systematically from a microscopic description is
still lacking. It is the purpose of the present paper to give one
step in this direction and provide an effective description of the
FQHE which is derived from a first principles microscopic action
describing interacting electrons in a strong external magnetic field
in a controlled way. In doing this, we are able to complement
previous descriptions and provide a way to take into account the
effects of a smooth boundary. These effects have been already
considered in \cite{horda,first
diff,Yang,Wan,plateau,exp-num,Chamon-Fradkin}, the main difference
with our approach is that our starting point is the microscopic
model and hence it does not rely on a phenomenological description.

The study of tunnelling of electrons into incompressible and
compressible quantum Hall states has been the subject of intense
research from both the theoretical and experimental sides (see
\cite{horda} and references therein). It is found that the
tunnelling conductivity is non-Ohmic, $I \propto V^\alpha$, with
$\alpha$ being a function of the filling fraction $\nu$. This
behavior can be understood within the chiral Luttinger liquid
description of the edge physics advanced by Wen \cite{Wen}.
Although tunnelling experiments in fractional quantum Hall effect
(FQHE) systems have shown certain degree of agreement with the
theoretical predictions obtained from the chiral Luttinger liquid
picture \cite{Wen}, there remain certain discrepancies which have
been addressed by different authors \cite{horda,first
diff,Yang,Wan,plateau,exp-num,Chamon-Fradkin}. In particular, a
reduction of the order of $10\%$ from the theoretical prediction
for the tunnelling exponent has been observed experimentally, as
well as the absence of the theoretically predicted plateaux
structure of this exponent \cite{first
diff,plateau,Chamon-Fradkin}. These issues have been the subject
of intense debate \cite{horda}.

Here we present an alternative derivation of the boundary effective
action, by including higher order terms in the low energy expansion
which are induced by the smoothness of the edge. 
%
%
This paper can be considered as a formal derivation of the model first presented in 
\cite{cort}.
In our calculations, the presence of a smooth boundary shows to be crucial
{\it e.\,g.} in the computation of tunnelling exponents. The effects
of the edge shape and the confining potential on these observables
have been recently addressed using numerical exact diagonalization
\cite{Wan}, where it was found that the result could deviate from
the chiral Luttinger liquid universal predictions. Our results are
in agreement with these findings.

In the simplest case of an almost sharp edge, the effective action
we obtain corresponds to  a self-interacting chiral boson. Using
this improved effective action we compute the tunnelling exponent
which, to one loop order, receives a small non-universal negative
correction depending on the electron density and the short distance
cutoff (the ``effective size" of the particles).

The paper is organized as follows: In Section 2, we present the
model which corresponds to a microscopic description of a fully
polarized self-interacting electron gas in a transverse magnetic
field. The action is invariant under relabeling of particles, a
symmetry that plays an important role in the continuum limit
(corresponding to the low energy description) as we show in Section
4. In Section 3 we rewrite the action by making a boost
transformation to eliminate the external electric field and we
discard terms quadratic in the particle velocities. This last
approximation is justified due to the presence of a strong external
magnetic field. We show that as a result of the interactions, the
system is projected into the fundamental state and that this
projection corresponds to the imposition of a constraint (Section
\ref{projectionII}) which defines the effective degrees of freedom
as the fluctuations along the flat directions of the potential. In
Section 4 we construct the low energy effective action by going to
the continuum limit and show that the resulting action is
topological. We show that the original symmetry under relabeling of
particles corresponds to the invariance under area preserving
diffeomorphisms of the effective action. A condition which arises
from the solution of the constraint is that our analysis is
applicable to incompressible states (Section \ref{flat}). Although
the constituent particles are fermions, the collective degrees of
freedom, which correspond to fluctuations along the flat directions
of the interaction potential, are described by bosonic fields.

In Section 5 we first solve the constraint by performing a gradient
expansion on the coordinate fields and then, by boosting back to the
original (laboratory) coordinate system, we arrive at one of our
main results, which is the effective action (\ref{aaa}). In Section
6 we look in more detail into the effective action close to a smooth
boundary, which requires to recover the discrete structure along the
direction perpendicular to the boundary, as expected. This leads
generically to a boundary effective theory of ${\cal N}$ coupled
chiral bosons, with ${\cal N} = W/a$, $a$ being the mean
inter-particle distance and $W$ the effective width of the edge
(\ref{c}). By integrating out the internal degrees of freedom, we
finally obtain the effective action for the boundary modes, which
corresponds to that of a chiral self-interacting bosonic field. From
this action, we compute the correction to the tunnelling exponent,
the result being in qualitative agreement with experiments. The
result obtained for the exponent is non-universal, which is a direct
consequence of the inclusion of a smooth edge \cite{nonuniv}. A
higher loop computation together with a renormalization group
analysis should be done, but this is out of the scope of the present
paper. The main aim here is to obtain for the first time an
effective field theory, starting from a microscopic description of
the real problem and which includes the effects of smooth boundary,
a feature present in almost all experiments. The observed edge
reconstruction effect \cite{exp-num} could be in principle recovered
within our approach by including the effects of the fluctuations
fields perpendicular to the flat directions of the interaction
potential (field $\Omega$ in (\ref{omega})). This is the subject of
future investigations. We conclude in Section 7 with a discussion of
the results and open perspectives.


\section{Polarized electron gas in a transverse magnetic field}

\subsection{The action}
\label{action-}

The most general action for a system of $N$ interacting electrons
is given by
\ba S = \int dt \left(\sum_p^N \frac m2{\dot\x_p}^{i\ 2} - V[\x_q^i,
\dot \x_q^i]\right) \ \ \ \ \ i=1,2 \ \ \ \ p,q=1,\cdots,N  \ , \ea
where the degrees of freedom of the system are represented by the
$2N$ functions $\x^i_p(t)$, {\em i.e.} the positions of the
electrons in a cartesian  coordinate system. These functions will be
our dynamical variables. For our purposes it is enough to consider
spinless fermions in two space dimensions since, as explained later, typical experiments are
performed in the presence of strong magnetic fields.

Note that this action is completely generic and makes no
simplification. The electrons interact with the background
and with each other by means of a general potential $V[\x_p,\dot
\x_p]$. This potential may depend simultaneously on the positions
and velocities of all particles ({\em i.e.} all $p=1,\cdots, N$),
and to keep this in mind we write its argument in square brackets.

We split the interaction into an external part and an
electron-electron interaction part $V[\x_p,\dot
\x_p]=V^{ext}[\x_p,\dot \x_p]+V^{int}[\x_p,\dot \x_p]$. Since
electrons interact independently with the external field, the
external part can be written as a sum of a single term per electron
$V^{ext}[\x_p,\dot \x_p] = \sum_p^N V^{ext}(\x_p,\dot \x_p)$ and we
then have
\ba S\! =\!\! \int \!dt\! \left(\sum_p^N \left(\frac
m2{\dot\x_p}^{i\ 2}-V^{ext}(\x_p^i,\dot \x_p^i)\right) -
V^{int}[\x_q^i,\dot \x_q^i] \right) \ , \ea
where the first term corresponds to the kinetic contribution.

Since the system is non-relativistic, the external potential
$V^{ext}(\x_p^i,\dot \x_p^i)$ can be expanded in powers of
velocities as
\be
V^{ext}(\x_p^i,\dot \x_p^i) = u(\x_p^i) + eA_j(\x_p^i) \dot \x_p^j +
\frac 12\delta m( \x^j_p){{\dot \x}^{i\ 2}_p}
+{\cal O}(\dot \x^{i\ 3}_p)
\label{Vext}
\ee
the first two terms representing the electric  and magnetic coupling
to the external field. The quadratic order in velocities represents
a position dependent correction to the mass that can be reabsorbed
in the kinetic term, while higher orders have been discarded.

On the other hand, the expansion of the inter-particle interaction
$V^{int}[\x_p^i,\dot \x_p^i]$ reads
\ba
V^{int}[\x_p^i,\dot\x_p^i] &=&
V^{int}[\x_p^i,0]
+\sum_q^N
\frac{
\,\partial V^{int}
}
{
\!\partial \dot \x_q^j
}
[\x_p^i,0]
\dot \x_q^j
+
\sum_q^N\sum_r^N
\frac{
\;\partial^2\!V^{int}
}
{
\partial \dot
\x_q^j \,\partial \dot \x_r^k } [\x_p^i,0] \,\dot \x_q^j\,\dot
\x_r^k +{\cal O}(\dot \x^{i\ 3})
\ea
The resulting terms can be
grouped according to the power of velocities involved, resulting in
a mass like (particle mixing) term, a magnetic term linear in
velocities, and an electric potential term
\ba
S\!\!&=& \!\!\!
\int \!dt \left(
\!\frac12\!\sum_{p,q}^N \left( \!(m+\delta m( \x^j_p))\delta_{pq}
\!-\!
\frac{\;\partial^2\!V^{int}}{\partial \dot \x_p^j \,\partial \dot
\x_q^k}[\x_r^i,0] \!\right)\dot \x_p^j\,\dot \x_q^k
\!-\!
\sum_p^N \left(\!eA_j(\x_p^i) \!+\! \frac{\,\partial V^{int}}{\!\partial \dot \x_p^j}[\x_q^i,0]
\!\right)\dot \x_p^j \!- \!\sum_p^N u(\x_p^i)
\!-\! V^{int}[\x_p^i,0]
 \right) \n
\ea
This is the action of our system. Up to this point no approximation
has been made (other than the non-relativistic limit)  and the
action has been kept completely general.

\subsection{\bf An important symmetry}

Since the particles are identical, the action is naturally
invariant under arbitrary permutations of the particle indices. We
will call this property {\em invariance under relabeling of the
particles}. This can be formally formulated as follows: the redefinition
\be p\ \to \ p' = p'(p) \label{relab1}\ee
acts in the dynamical variables as
\be \x_p^i= \x^i_{p(p')} \equiv {\x^i}'_{p'} \label{relab2}\ee
leaving the action unchanged.

\section{\bf The interactions and the resulting projection into the fundamental state}
\label{projection}

The general expansion for the external potential
(\ref{Vext}) resulted in an electromagnetic interaction in three
space-time dimensions. We define the corresponding external magnetic
field as $\vec B = B \check{z}$, with $B=\partial_i
A_j\epsilon^{ij}$ and we assume it is constant and homogeneous in
all space to conform with the typical experimental setup. Moreover,
we take it to be very strong so as to completely polarize the
particle spin in the $\check z$ direction, justifying in this way
our choice of spinless fermions to represent the electrons. The
electric field on the other hand is defined as $E_i=-\partial_iu$.
It is assumed to be zero in the interior of the region of space
occupied by the sample, and smoothly growing at the edge of that
region.

\subsection{A convenient coordinate system}
\label{boostsect}

In this setup, external electric and magnetic fields are
perpendicular and, in the absence of additional interactions,
we can discard the acceleration term in the equations of motion
$\dot v_i = E_i+\epsilon_{ij}v_j B\simeq0$. Solving for $v_i$ we see that
the resulting classical particle motion takes place in the direction normal $E_i$.
With an appropriate boost, we could jump to a coordinate system in which the particle is at
rest and there is no electric field. Even in the presence of an
inter-particle interaction as in our case, we are interested in
eliminating the external electric field by such a boost. The
corresponding velocity will be
\be v_i = -\frac1{B}\,\epsilon_{ij}\partial_ju \label{veloc}\ee
%
Using the invariant expression $B^2(1-(\partial_iu)^2/B^2)$, and
expanding in powers of $E/B$, we can solve for the magnetic field in
the boosted coordinate system $\tilde B$ as
\be \tilde B \simeq B\left(1- \frac{(\partial_iu)^2}{2B^2}\right)
\ee
%
%
%

The action in the new coordinate
system reads
\ba S\!\!&=& \!\!\! \int \!dt \left( \!\frac12\!\sum_{p,q}^N \left(
\!(m+\delta m( \tilde \x^j_p))\delta_{pq} \!-\!
\frac{\;\partial^2\!V^{int}}{\partial \dot {\tilde \x}_p^j
\,\partial \dot {\tilde \x}_q^k} \!\right)\dot {\tilde \x}_p^j\,\dot
{\tilde \x}_q^k \!-\! \sum_p^N \left(\!e\tilde A_j(\tilde \x_p^i)
\!+\! \frac{\,\partial V^{int}}{\!\partial \dot {\tilde \x}_p^j}
\!\right)\dot {\tilde \x}_p^j \!-\! V^{int}
 \right)\ \
\ea
where $\tilde A_j(\tilde \x_p^i)$ is the external vector potential
in the new coordinates. Note that in this frame $\tilde
u(\tilde\x_p^i)=0$ and the external magnetic field $\tilde B(\tilde
\x_p^i)$ is non-homogeneous. In other words, the effects of the
external electric field in the old coordinate system $\x$, is now
encoded in $\tilde B(\tilde \x_p^i)$ in the new system $\tilde
\x_p^i$. This is very important since, as we will see, it turns out
to lead to localization of the excitations.

On the other hand, it is evident that the relabeling symmetry
persists in the new variables, taking the same form as before, {\em
i.e.}
\be \tilde \x_p^i  = {\tilde {\x'}}_{p'}^i \label{relab3}\ee

\subsection{The projection constraint and a new form of the action}
\label{projectionII}

Due to the presence of a strong external magnetic field, we can
discard the kinetic terms quadratic in velocities, as well as the
magnetic inter-particle interaction ${\,\partial
V^{int}}\!/{\partial \dot {\tilde \x}_p^j}$. In a perturbative
quantization approach, this approximation corresponds to the
projection into the lowest Landau level (LLL). The scrupulous reader
can keep all the terms, then go through a quantization procedure and
project the result into the LLL.

The resulting effective Lagrangian in then linear in velocities and
the resulting action for the degrees of freedom in the ground state
is
\ba S&=& \int \!dt \left(\sum_p^N  e\tilde A_j(\tilde \x_p^i)\dot
{\tilde \x}_p^j
- V^{int}[\tilde \x_p^i] \right)\ ,
\ea
Where we called $V^{int}[\tilde \x_p^i]=V^{int}[\x_p^i,0]$. To be
consistent, we should supplement this action with the condition
enforcing the low energy projection, namely that the system remains
in its ground state ${\cal H}= E_{min}$. Within our approximation,
this can be written as
\be
V^{int}[\tilde \x_p^i]
=0
\label{fundamental}
\ee
where we have eliminated any non-vanishing contribution to the
ground state energy by a shift in the potential. 
This constraint can be enforced at the action level by introducing a
Lagrange multiplier $\lambda(t)$ to obtain

\ba S&=& \int \!dt \left(\sum_p^N  e\tilde A_j(\tilde \x_p^i)\dot
{\tilde \x}_p^j  - \lambda V^{int}[\tilde \x_p^i] \right)\ ,
\label{cont-act}\ea
the equations of motion being now obtained  by varying the action
with respect to $\tilde \x_p^i$ and the Lagrange multiplier
$\lambda(t)$. This new form of the action is still invariant under
the relabeling symmetry.

\section{\bf The continuum limit and the resulting topological field theory}
\label{continuum}

We assume that all the important physical scales involved in our
problem are large compared to the inter-particle distance. These
scales can be characterized in terms of the relevant length of the
external field $\tilde B/\partial_j \!\tilde B$. Since the
derivatives involved in this definition can differ significatively
when taken in different directions, so does the accuracy of this
assumption. In the present section we assume that it is valid in any
direction, an assumption which is well justified in the bulk. We
will come back to this point in the case in which this assumption
ceases to be valid, {\it i.e.} when analyzing boundary effects in
Section \ref{slicing}.

\subsection{The continuum space}

We take the continuum limit in the above system as follows: first we
define our continuum space and functions living on it by the
following steps:
\begin{enumerate}

\item We introduce an auxiliary two-dimensional space parameterized
by variables $\tilde y^i$.

\item We set  a (not necessarily regular) lattice $\tilde
y_p^i$ in that space by the condition
 $\tilde y^i_p=\tilde \x^i_p(0)$, that will be useful later.

\item We use the discrete quantities $\tilde\x^i_p(t)$ of our
problem to define lattice functions $\tilde\x^i(\tilde
y_p^j,t)$ on the auxiliary $\tilde y_p$ space by means of
\ba \tilde\x^i\!(\tilde y_p^j,t) &=& \tilde \x_p^i(t) \label{culo}
\ea
In other words, we label each particle $p$ with their initial
position $\tilde y_p^i$.
\item We re-interpret these lattice quantities as the values at
the lattice points of some continuum fields $\tilde \x^i(\tilde
y^j,t)$
\be \tilde \x^i(\tilde y_p^j,t) = \left.\tilde \x^i(\tilde
y^j,t)\right|_{y^i=y_p^i} \ee
These continuum fields can be understood as an interpolation of the
discrete quantities representing the particle positions at a given
time in the intersticial region of $\tilde y$-space. There are many
interpolation methods that could be used for that \cite{NR},
however, we do not need an explicit prescription for it. Note that
this re-interpretation is consistent with the initial condition on
the continuum variables
\be \tilde \x^i(\tilde y^j,t=0)\vert_{\tilde y^j = \tilde y_p^j} =
\x^i(\tilde y_p^j,0)=\tilde \x^i_p(0)=\tilde y^i_p \ee

\end{enumerate}

\subsection{The resulting field theory}
We reformulate now the dynamics in terms of the continuum variables
introduced in (\ref{culo}) by starting from the discrete action
(\ref{cont-act}). The procedure goes as follows:
\begin{enumerate}

\item We make a simplicial decomposition of the $\tilde y^{i}$ plane
consistent with the lattice, {\em i.e.} having a single point
$\tilde y^i_p$ for each simplex. It is easy to convince oneself that
this is always possible and that in two dimensions the simplices are
triangles that can be defined as having a lattice point at each
vertex. The easiest choice is a simplicial decomposition
topologically equivalent to the triangular lattice.

%

\item We call $\Delta \tilde A_p$ the (small) area of the $p$-th simplex and multiply
and divide by it the $p$-th term in the action (\ref{cont-act}).
This results in a Riemann approximation for the integral
\ba \sum_p^N  e\tilde A_j(\tilde \x(\tilde y_p^i,t))\partial_t
{\tilde \x}(\tilde y_p^j,t) &=& \sum_p^N \Delta \tilde A_p
\frac1{\Delta \tilde A_p} \, e\tilde A_j(\tilde \x(\tilde
y_p^i,t))\partial_t {\tilde \x}(\tilde y_p^j,t) = \n &\simeq& \int
d^2\tilde y \,\tilde \rho(\tilde y)\, e\tilde A_j(\tilde \x(\tilde
y^i,t))\partial_t {\tilde \x}(\tilde y^j,t)+ {\cal O}\left[\sum_{p
}\!\left|\partial_i \left(e \tilde \rho \tilde A_j(\tilde \x
)\partial_t {\tilde \x}\right)\Delta y^i_p\right|_{(\tilde
y_p,t)}\!\!\!\!\!\!\Delta \tilde A_p\right] \label{riemann} \ea
where the {\em number density} function $\tilde \rho(\tilde y)$
stands for the interpolation of $\tilde \rho(\tilde y_p)\equiv
1/\Delta \tilde A_p$. This approximation of the sum remains accurate
as long as the jump on the argument when going from one lattice
point to another stays small. We have kept terms up to linear order
in these jumps and provided an estimation of the error made in that
process that will be useful later.
\end{enumerate}

Applying the above steps explicitly we obtain the continuum low
energy effective theory
\ba S&=& \int \!dt \left(\int d^2\tilde y \tilde\rho\, e\tilde A_j(\tilde \x^i )
\,\partial_t{\tilde \x } - \lambda V^{int}[\tilde \x ] \right)\ ,
 \label{action2} \ea
in terms of the continuum fields $\tilde \x^i(\tilde y,t)$ and the
Lagrange multiplier $\lambda(t)$.

We are interested in solutions $\tilde \x^i(\tilde y,t)$ of the
field theory (\ref{action2}) that are good approximations of the
true solutions $\tilde \x_p^i(t)$ of the discrete microscopic system. In
particular this means that the error estimated in (\ref{riemann})
has to be small when evaluated on them. In other words, we should
discard those solutions whose derivatives are large compared to the
inter-particle distance.

Note that the action contains a local velocity-dependent term,
coming from the external magnetic field, plus a non-local potential
constraint, coming from the inter-particle interaction. The
topological nature of the resulting field theory is evident in view
of the absence of any spatial derivatives.

In this part of the calculation a new function $\tilde \rho (\tilde
y)$ has appeared. We need some suitable ansatz for the form of this
function. Since is represents the particle number density in the
ground state, a judicious ansatz would be to choose it consistently
with the forces on these particles in that state. In the $\tilde y$
plane, particles at rest experience no forces other than their
repulsive interaction, so we can safely choose $\tilde \rho(\tilde
y)$ as a constant $\tilde \rho_o$ in this plane.

\subsection{Symmetry under Area Preserving Diffeomorphisms and gauge invariance}

The relabeling symmetry (\ref{relab1})-(\ref{relab2}),
(\ref{relab3}) is inherited by the continuum limit theory, where it
manifests as a symmetry under area preserving diffeomorphisms.
To see this, we first note that the area $\Delta\tilde A_p$ of the
simplex associated to the particle $p$ becomes, after a relabeling,
\be \Delta \tilde A_p = \Delta \tilde A_{p(p')} = \Delta \tilde
A'_{p'} \label{area}\ee
The second equality is evident from the fact that $\Delta \tilde
A_p$ becomes, after relabeling, the area of the simplex associated
to the particle $p'$.

In the continuum limit, the symmetry can be written as
\be y \ \to \ \tilde y' = \tilde y'(\tilde y) \ee
while the fields transform according to
\be {\x^i}'(y',t)= \x^i(y,t)\ee

The equality (\ref{area}) then implies that
\ba d^2\tilde y = d^2\tilde y' \ea
Then the transformation corresponding to the relabeling of particles
becomes an area preserving diffeomorphism in the continuum theory

Due to the definition of $\tilde \rho$, the combination $\tilde \rho
\,d^2\tilde y$ is invariant under {\it any} transformation. Then the
kinetic term in the action is obviously invariant under area
preserving diffeomorphisms. The potential, on the other hand,
inherits its invariance from the discrete case.

We conclude that our system has a continuos invariance. In the case
in which $\tilde B$ is constant, this corresponds to the gauge
invariance of the usual Chern-Simons description \cite{Wen}. On the
other hand, in the case in which $\tilde B$ is space dependent, the
only transformations leaving the action invariant are those which
are time independent. This implies that gauge invariance is broken
in our theory, in a similar way as the imposition of a boundary
breaks gauge invariance in \cite{Wen}, enforcing the introduction of
the boundary chiral boson. In our formulation this chiral boson
theory arises naturally as a consequence of the low energy
constraint. This is the first insight of one of the main claims of
this paper, {\em i.e.} that a smooth boundary, represented by an
external potential varying in space, can be described by a theory
whose qualitative behavior is similar of that of the chiral boson.

\subsection{Flat directions of the ground state and a simpler form of the constraint}
\label{flat}
The Lagrange multiplier in the action imposes the constraint that
forces the system to be in the ground state. To obtain the remaining
degrees of freedom we would then need to solve this constraint. This
seems impossible without knowing explicitly the form of the
potential, but we show in this section that this is indeed
simplified by the symmetry under area preserving diffeomorphisms of
our system. In consequence we can replace the constraint for an
explicit one, without giving any details on the potential.

The condition (\ref{fundamental}) fixes all directions except
those that leave the potential invariant. In other words, particles can
move along the flat directions of the potential, while staying in
the ground state. These directions correspond then to the
remaining degrees of freedom.
To determine them, let us note that any small deformation  $\tilde
\x^i(\tilde y,t)=\tilde \x^i_o(\tilde y,t)+\delta\tilde \x^i(\tilde
y,t)$ around a given minimum of the potential $\tilde \x^i_o(\tilde
y,t)$, can be written in the form
\be \delta\tilde \x^i(\tilde y,t) = M_{ij}(\tilde y,t)
\left(\partial_j\Omega(\tilde y,t) +
\epsilon_{jk}\partial_k\Lambda(\tilde y,t)\right) \ee
for suitable $(\Omega,\Lambda)$, when $M$ is a given invertible
matrix field that we choose for convenience as
$M_{ij} = \partial_i \tilde \x^j_o(\tilde y,t)$.
With this choice  we can write
\ba \delta \tilde \x^i(\tilde y,t) &=&
\partial_i \tilde \x^j_o(\tilde y,t) \left(\partial_j\Omega(\tilde y,t)
+ \epsilon_{jk}\partial_k\Lambda(\tilde y,t)\right) \ea
or in other words
\ba \tilde \x^i(\tilde y,t)&=& \tilde \x_o^i(\tilde y_j +
\partial_j\Omega(\tilde y,t)+ \epsilon_{jk}\partial_k\Lambda,t) \label{omega}\ea
implying that any small deformation of the configuration $\tilde
\x^j_o(\tilde y,t)$ can be obtained by a small diffeomorphism on
$\tilde y$. The $\Omega$ part represents dilatations of the $\tilde y$ plane,
while the $\Lambda$ part corresponds to area preserving deformations.
Going now to the potential, this implies
\ba V[\tilde \x_o(\tilde y,t)+\delta \tilde x^j (\tilde y,t)] &=&
V\left[\tilde\x_o(\tilde y_j+
\partial_j\Omega + \epsilon_{jk}\partial_k\Lambda,t)\right]
\n &=& V\left[\tilde \x_o(\tilde y_j+ \partial_j\Omega(\tilde
y,t),t)\right] \label{degrees}\ea
where in the last equality we use the invariance of the potential
under the area preserving diffeomorphism generated by $\Lambda$,
$\delta\tilde y_i = \epsilon_{ij}\partial_j\Lambda$. We conclude
that the condition (\ref{fundamental}) fixes $\Omega$ and leaves
$\Lambda$ completely free. The flat directions of the potential are
then the variations of $\tilde \x^i_o(\tilde y,t)$ generated by area
preserving diffeomorphisms.

In particular,  since $\tilde \x^i(\tilde y,0)=\tilde y^i$, we
conclude that the time evolution map, going from $\tilde y^i$ to
$\tilde\x^i(\tilde y, t)$, is area preserving. In other words, the Jacobian of the time evolution is unity, implying that a
physical configuration satisfies
\be |\partial_i\tilde \x^j|=1 \label{constraint} \ee
This constraint has a simple and intuitive interpretation: note that
the particle density (in $\tilde \x^i$ space) at time $t$ is given
in terms of the initial density $\tilde \rho_o$ by the formula
\be
\tilde \rho(t,\tilde \x) = \frac{\tilde \rho_o}{|\partial_i \tilde \x^j|}\equiv\tilde \rho_o
\label{incompressible}
\ee
{\em i.e.} our evolution is that of an incompressible fluid. For
this reason, we will concentrate in what follows in incompressible
states. To describe compressible states on the other hand, we would
have to consider the $\Omega$ degree of freedom.

It should be clear now that we can replace the constraint
(\ref{fundamental}) by the more explicit one (\ref{constraint}). In
other words, we can replace (\ref{action2}) by the completely local
action
\ba S&=& \int \!dt \int d^2\tilde y \tilde\rho_o\left( e\tilde A_j(\tilde \x^i )
\,\partial_t{\tilde \x } - \lambda\left(|\partial_i\tilde \x^j|-1\right)\right)
\label{action.local} \ea
where now $\lambda(\tilde y,t)$ is a local Lagrange multiplier
imposing (\ref{constraint}).

As a conclusion, in this section we have been able to use the symmetries
to describe the low energy dynamics independently of the explicit
form of the inter-particle potential.

\subsection{The conserved charge and an immediate consequence}

Since the action is invariant under the group of area preserving
diffeomorphisms in the $\tilde y^i$ plane, we can calculate the
corresponding N\oe ther charge. To do that, we write the
infinitesimal transformation as
\be y_i'= y_i +\epsilon_{ij}\partial_j \Lambda \ee
and the charge reads
\be Q = \frac{e}2\int d^2y\,\tilde B(\tilde
\x)\epsilon_{ab}\epsilon^{ij}\partial_j\tilde \x^b\partial_i \tilde
\x^a \Lambda \ee
Moreover, since this charge is conserved for any function $\Lambda$,
its integrand is conserved, {\it i.e.} the magnitude
\be \tilde B(\tilde \x) \epsilon_{ab}\epsilon^{ij}\partial_j\tilde
\x^b
\partial_i\tilde \x^a = 2 \tilde B(\tilde
\x)|\partial_i \tilde\x^j| = 2\tilde B(\tilde \x) \label{contr} \ee
%
is conserved. Here in the last equality we used the constraint
(\ref{constraint}). We can fix its value according to the initial
condition as
\be  \tilde B(\tilde \x) = \left.\tilde B(\tilde \x)\right
|_{t=0}=\tilde B(\tilde y)\ee
Note that this relation implies that the value of the magnetic field
at the particle position  is the same along all its motion. In other
words, the particle moves along the level lines of the function
$\tilde B$. Again, this may be interpreted as a behavior analogous
to that of the Chern-Simons theory, in which excitations move along
the boundary, and can be described by a chiral boson. Since in our
case the boundary has been replaced by a smooth space-dependent
external field, this translates into a motion along the level lines
of this external field, and the corresponding chiral boson theory
will necessarily have to take this into account.

\section{\bf Solution of the constraint and the resulting boundary theory}

\subsection{Solution of the constraint as a gradient expansion}

The next step is to solve the constraint (\ref{constraint}) and
rewrite the action in terms of the dynamical degrees of freedom.
Here and in what follows, we use as parameters the filling fraction
in the $\tilde \x$ plane, defined as $\tilde \nu = \tilde \rho/e
\tilde B$, and the effective (dimensionfull)
parameter $\theta = 1/(2\pi\tilde\rho_o)$ defined in \cite{Susskind} as the
square of the non-commutativity length.

If we parameterize the $\tilde \x^i$ field with the help of a
slowly varying bosonic field $\phi$ according to the following
gradient expansion
\be \tilde \x^k=\tilde y^k + \theta\,{\epsilon^{kl}}\partial_l\phi +
\frac {\theta^2}2\,
\epsilon^{kl}\epsilon^{ij}\,\partial_i\phi\,\partial_j\partial_l\phi
+{\cal O}(\theta^3) \label{solution} \ee
then (\ref{constraint}) is solved up to order $\theta^3$. This field
represents the effective degrees of freedom after projection into
the ground state. It can be interpreted to first order in $\theta$
as representing the coordinate fluctuations. Incidentally we see in
(\ref{solution}) that $\phi$ is defined up to the addition of an
arbitrary function of time $\phi\to\phi+g(t)$, which is the usual gauge invariance of the chiral boson theory.

Replacing this solution in the action (\ref{action.local}) we get
\ba S &=& \frac{1}{8\pi^2} \int  \!dt d^2\tilde y\,\tilde
\nu^{-1}{\epsilon^{ij}}\!\left(
\partial_{j}\phantom{{x^2}^2}\!\!\!\!\!\!\!\!\!\!
\left(\phi\,\partial_{i}\partial_t \phi\right)+\frac \theta 3\partial_{j}\!
\left(\epsilon^{k{l}}\partial_{k}
\partial_t \phi\,\partial_i\phi\,\partial_{l}\phi
\right) \phantom{{x^2}^2}\!\!\!\!\!\!\!\!\!\!\right)\,+\,{\cal
O}(\theta^2) \label{action.local.solved} \ea

\subsection{\bf Boost back}

Note that the variables $\tilde \x^i(\tilde y,t)$ refer to the
position of the particle initially located at $\tilde y$. This is
the Lagrangian description of a fluid, where each fluid element is
labeled with its initial position and then followed along its motion
throughout the plane. But the initial position $\tilde y^i$ is given
in an awkward coordinate system that is related to that of the
laboratory by the transformation discussed in Section
\ref{boostsect}. We would like to refer things to the laboratory
coordinate system by going back on the original  frame as $y^i =
y^i(\tilde y)$. Note that for convenience we do not apply this
operation to the $\tilde \x^i$ fields. They still correspond to the
position at time $t$ in the tilted system, but now the fluid
element is identified by its initial position $y^i$ on the laboratory
frame.

Under this transformation, the action becomes
\ba \frac{1}{8\pi^2} \int  \!dt d^2 y\,\nu
^{-1}\,{\epsilon^{ij}}\!\left(
\partial_{j}\phantom{{x^2}^2}\!\!\!\!\!\!\!\!\!\!
\left(\phi\,\partial_{i}(\partial_t-v^a\partial_a) \phi\right)+\frac
\theta 3\partial_{j}\! \left(\epsilon^{k{l}}\partial_{k}
(\partial_t-v^a\partial_a) \phi\,\partial_i\phi\,\partial_{l}\phi
\right) \phantom{{x^2}^2}\!\!\!\!\!\!\!\!\!\!\right)
\label{action.local.solved.boosted} \ea
where we defined the position dependent filling fraction in the
(physical) $y$ plane as $\nu = \rho/eB$, and it satisfies
\be \nu = \frac{\tilde \rho_o}{eB}\left(1+\frac
{(\partial_iu)^2}{2B^2}\right) \ee
On the other hand, the velocity of the excitations is defined as in
eq.(\ref{veloc})

In the special case in which $B$ is constant the action is a total
derivative. The usual procedure is to assume this and then impose a
boundary to the region in which the fluid moves. We choose a
different way here, allowing the parameter $\nu$ to change in space
and introducing through it the information about the geometry,
without adding any boundary. A further integration by parts gives
\ba S &=& - \frac{1}{8\pi^2} \int  dt d^2\!y
\,t^i\!\!\left(\phantom{{x^2}^2}\!\!\!\!\!\!\!\!\! \,
\phi\,\partial_{i}(\partial_t-v^a\partial_a)\phi + \frac{\theta}{3}
\epsilon^{bc}\,
\partial_{b}(\partial_t-v^a\partial_a)
\phi\,\partial_i\phi\,\partial_{c}\phi
\phantom{{x^2}^2}\!\!\!\!\!\!\!\!\!\!\right) \label{aaa} \ea
where we defined the vector field $t_i(y)$, tangent to the level line of $\nu$ at $y$, as
\ba t^i &=&\epsilon^{ij}\partial_j (\nu^{-1})\ea
We see that the degrees of freedom are localized at the positions at
which the change of variables discussed in Section \ref{boostsect}
is non-trivial, {\em i.e.} where the external electric field
$\partial_iu$ is non-constant. Moreover they propagate in the
direction of $t_i$ along which  $E^2$ is constant.

Here we see that, wherever the sample is homogeneous, the
derivatives in front of each term vanish, leaving us without any
dynamics. On the other hand, the dynamical degrees of freedom
localize at places where there is a change in the properties of the
material ({\em i.e.} the filling fraction). This is precisely what
happens at the edge of the sample, and it is at the core of our
argument.

To stress this point, let us suppose that the filling fraction
changes as a step function, then their derivatives will provide
delta functions leading us to a one dimensional boundary theory,
which corresponds to a Chiral Luttinger Liquid description
\cite{Wen}. Our treatment however is more general in the sense that
it includes the possibility of dealing with smooth edges.

The action (\ref{aaa}) is written completely in terms of the
dynamical degrees of freedom $\phi$. Note that this is not a
boundary theory because it is defined in the full two dimensional
space. Nevertheless, the degrees of freedom are bounded to the
region  in which there is a change in the parameter $\nu$ and, as
predicted, propagate chirally along their level surfaces.

\section{Recovering the discrete structure near the edge}
\label{slicing}

When we approach the edge, the derivatives of the functions entering
into the action are no longer negligible. Then the assumption we
made when we took the continuum limit (\ref{riemann}) is not
satisfied, and the solutions of the continuum theory do not
represent faithfully the underlying discrete system. To go around
this problem, we need to recover the discrete structure near the
edge. Calling $y_\parallel$ the direction that runs along the edge
and $y_\perp$ the perpendicular direction, we see that the $y_\perp$
derivatives become big in the edge region.
 Since $y_\parallel$ derivatives are still small, we can keep the $y_\parallel$ integral while in
the direction $y_\perp$ transverse to the edge, we proceed as
follows:

\begin{enumerate}

\item Cut the space in the $y_\perp$ direction in ${\cal N}$ slices at $y_\perp^n$ of
width $\Delta y_\perp \sim a$.

\item Replace the integral by a sum over the values of the
integrand evaluated at each slice, and define
$\phi_n(y_\parallel)\equiv\phi(y_\parallel,y_\perp^n)$.

\item Replace all the $y_\perp$ derivatives by its finite difference
approximation $$\partial_\perp \phi(y_\parallel,y_\perp^n) =
(\phi(y_\parallel,y_\perp^n)-\phi(y_\parallel,y_\perp^{n-1}))/\Delta
{y_\perp}$$
\end{enumerate}

With this method we obtain an effective theory for the fields
$\phi_n(y_\parallel)$ living on each slice $n$, with action
\ba S&=&-\frac{1}{8\pi^2}  \int  \!dt \,dy_\parallel
\sum_{n=0}^{\cal N}
\left(\,\phantom{{y^2}^2}\!\!\!\!\!\!\!\!\!\!\kappa_n {\cal
L_F}(\phi_n) +  \kappa'_n \left({\cal L_S}(\phi_n) + {\cal
L_I}(\phi_n,\phi_{n-1}) \phantom{f^F}\!\!\!\!\!\!\!\right)
\phantom{\frac12}\!\!\!\!\!\right) \label{c} \ea
where we have defined the free, self interaction and interaction
Lagrangians as
\ba {\cal L_F}(\phi_n)&=&
\phi_n\,\partial_\parallel(\partial_t-v_n\partial_\parallel)\phi_n
\n {\cal L_S}(\phi_n) &=&
\phi_n(\partial_t-v_n\partial_\parallel)(\partial_\parallel\phi_n)^2
\n {\cal L_I}(\phi_n,\phi_{n-1}) &=&
-\phi_{n-1}(\partial_t-v_n\partial_\parallel)(\partial_\parallel\phi_n)^2
\ea
and the constants $\kappa_n$ and $\kappa'_n$ are given by
\be \kappa_n =\frac 1{\nu_n}-\frac 1{\nu_{n-1}} \ , \ \ \ \ \ \ \
\kappa'_n =\frac \theta{2\Delta y_\perp}\left(\frac {1}{\nu_n}-
\frac{1}{\nu_{n-1}}\right) \ . \label{levels} \ee
It is important to note that the field $\phi_n$ will enter into the
action only when the constants $\kappa_n, \kappa'_n$ are
non-vanishing, {\it i.e.} if there is a change in the properties of
the sample between the slices $n$ and $n+1$.

The gauge invariance is now $\phi_n\to\phi_n+g(t)$ adding the same
$g(t)$ to all the $\phi$'s.

Note that when the sample has a sharp edge, {\it i.e.} if the
density changes within a region whose width $W$ is smaller that the
slicing length $\Delta y_\perp$, {\it i.e.} $W<\Delta y_\perp \sim
a$, then the whole procedure is not applicable and the boundary
theory corresponds to the usual chiral boson theory \cite{Wen}.

Let us suppose that the edge is wider than the slicing length, so
that we have to keep a finite number of terms of the sum in
(\ref{c}), $n=1,\cdots , N$.

We have then a single non-dynamical field $\psi\equiv \phi_{-1}$ in
the interaction term ${\cal L_I}(\phi_0,\phi_{-1})$, all other
fields appearing in the ${\cal L_I}(\phi_n,\phi_{n-1})$ terms being
dynamical. The integration of this multiplier field will enforce a
constraint on the field $\phi_0$ to which it is coupled. When
solved, $\phi_0 = f(y_\parallel+v_0t)$, and replaced in the action,
it implies that the interaction term with $\phi_1$ takes the form
\ba
 -\frac{1}{8\pi^2} \kappa'_1
\int  \!dt \,dy_\parallel
f(y_\parallel+v_0t)(\partial_t-v_1\partial_\parallel)(\partial_\parallel\phi_1)^2
\label{cc} \ea
which is different from zero provided that the two velocities are
not equal. Since we have assumed that the electric field in the edge
zone is approximately constant in regions of size $a\sim\ \Delta
y_\perp$, hence $v_0 \approx v_1$ and we can discard this term
within the present approximation.

We have then obtained an action very similar to the original one but
without any Lagrange multiplier
\ba S= -\frac{1}{8\pi^2} \int \!dt \,dy_\parallel \sum_{n=1}^{\cal
N}\left(\phantom{{y^2}^2}\!\!\!\!\!\!\!\!\! \kappa_n {\cal
L_F}(\phi_n) + \kappa'_n \left({\cal L_S}(\phi_n) +
(\delta_{n1}-1){\cal L_I}(\phi_n,\phi_{n-1})
\phantom{2^2}\!\!\!\!\!\right)\phantom{\frac12}\!\!\!\!\!\!\right)
\label{cb} \ea
where the lower bound of the sum has changed and the coefficient
$(\delta_{n1}-1)$ ensures that there is no interaction term for
the field $\phi_{1}$.

\section{An application of the model: tunneling exponents}
\label{tuneling}

Let us consider the case of an almost sharp boundary, in which the
density can be well approximated by choosing ${\cal N}=1$. In that
case the action is
\ba S = -\frac{1}{8\pi^2} \int  \!dt \,dy_\parallel \left(
\phantom{{y^2}^2}\!\!\!\!\!\!\!\!\!\kappa
\,\phi\,\partial_\parallel(\partial_t-v\partial_\parallel)\phi +
\kappa'
\phi\,(\partial_t-v\partial_\parallel)(\partial_\parallel\phi)^2
\phantom{{y^2}^2}\!\!\!\!\!\!\!\!\!\right) \label{action} \ea
where we have dropped the subindexes since we have a single field.
This is a chiral boson theory with a cubic higher derivative self
interaction, the latter arising essentially from the smoothness of
the edge.
To see the effect of this last term in the physical properties, we
compute the propagator since it is directly related to the
tunnelling exponent \cite{Wen}.

In Fourier space we write for the Feynman propagator
\be G(p,\omega_p) = \frac {2\pi} {\kappa
p\,(\omega_p-vp)+i\epsilon} \label{propa} \ee
while the vertex is proportional to
\ba -\kappa'\delta(p+q+r)(\omega_p-vp)\,q\,r \ea
Then, the one loop corrected propagator reads
\ba
G'(p,\omega_p) &=&\frac {2\pi} {\kappa p\,(\omega_p-vp)-\delta
G(p,\omega_p)}
\ea
We consider those Feynmann diagrams whose contribution to the
corrected propagator is
\ba
\delta G(p,\omega_p) =\pi\Lambda^2\frac{\kappa'^2}{\kappa^2 }
(\omega_p-v p) p
\ea
where $\Lambda$ is a momentum cutoff. Replaced in (\ref{propa})
this gives

\be
G(p,\omega_p) = \frac 1 {(\kappa +\delta
\kappa)p\,(\omega_p-v p)}
\ee
where the non-universal correction to the level is given by

\ba
\delta\kappa &=& -\pi \frac{\kappa'^2}{\kappa^2}\Lambda^2
\ea
Note that the effective coupling is
${\kappa'}_n^2/\kappa_n^2=\theta^2/{4\Delta x^2_\perp}$, being
independent of the details of the boundary. A non vanishing
correction to the velocity arises from the remaining diagrams but it
produces no physical consequences.

Since the tunnelling exponent is directly related to the level by
$\alpha\approx\kappa$, the appearance of a correction for the
latter implies, to first order, a correction to the tunneling
exponent. It should be stressed at this point that the correction is negative, in qualitative accordance
with the experimental results. This is the main result of this section.

To estimate the magnitude of this correction, we need to relate
the momentum cutoff $\Lambda$ to the minimal space distance
measurable $a$, which naturally leads to the identification
$\Lambda = \pi/a$. Using (\ref{levels}) we then get for this
choice of the cutoff

\be \delta\kappa =  - \frac{\pi}{(4a^2\tilde\rho_o)^2} =
-\frac{\pi}{16} \approx -0.196 \ee
where we have further identified $\tilde \rho_o=1/a^2$. The
predicted dependence on the density of this non-universal correction
could in principle be tested experimentally.

Putting it all together this leads to a linear dependence of the
tunneling exponent $\alpha$ as a function of $1/\nu$
%
which is very close to the experimental fit presented in \cite{first
diff} (see for example Fig.\ 3 of this reference). However it should be
kept in mind that our present result is valid for incompressible situations.
It can be observed that the departure of our result from the experimental data
becomes greater for increasing $1/\nu$, which could be attributed to
the need to work with ${\cal N}>1$ in (\ref{cb}). This could in turn
be related to the need to consider a wider boundary region. The
analysis of the higher ${\cal N}$ case will be presented separately
\cite{CG}.

\section{Conclusions}

To summarize, we have constructed an effective field theory
that describes the boundary region of a fractional quantum Hall effect droplet, starting from the
microscopic description. We included the interplay between the
granularity of the fluid (the inter-particle distance $a$) and the
effects of a smooth boundary. This extended boundary has been
defined by means of a non-vanishing electric field, that leads to a
space-dependent filling fraction within this region. The physical
excitations of the system are localized at the boundary, and are
described by a system of self interacting chiral bosons.

All the approximations we made along our calculations are under
control: by taking the non-relativistic limit in the strong magnetic
field regime, we enforced the projection into the LLL. Keeping the
interaction between particles completely generic, this low energy
projection results in a constraint on the dynamical degrees of
freedom of the model. With the help of the symmetries of the system,
we were able to replace this constraint by an equivalent one that
implies incompressibility and is completely independent of the
details of the interaction potential.

As an example of the possible applications of our approach, we have
computed the correction to one loop order of the tunnelling exponent
to lowest order in the boundary width, which is in qualitative
agreement with experimental results.

The results presented here can in principle be improved in a
systematic way. Wider boundary regions can be easily treated within
the present approach, by simply taking higher values of ${\cal N}$
in (\ref{cb}). Higher loop computations could also be envisaged,
however they would involve the solution of the constraint to higher
orders in $\theta$. This is out of the scope of the present paper.
As a further improvement, one can describe compressible situations
by including the effects of the degrees of freedom transverse to the
flat directions of the potential (described by $\Omega$ in
(\ref{degrees})).

Other approaches to include a boundary in a Chern-Simons
description of a quantum Hall droplet have been proposed
\cite{Poly}. In this context, it would be interesting to study their
connection to our model.

It should be mentioned that the experimental data in \cite{first
diff} was reanalyzed in \cite{plateau} in view of the results of
\cite{Chamon-Fradkin}, and certain degree of agreement with the
plateau structure was obtained. Here we propose an alternative
description which provides a reasonable agreement with the raw
experimental data as presented in \cite{first diff}. It would be
interesting to study the higher order theory that we have
constructed in the present paper along the lines of the approach
presented in ref.\ \cite{Chamon-Fradkin}. This could provide a
closer agreement between theoretical and experimental results.

It would be also interesting to analyze the consequences of the
higher order corrections induced by the smoothness of the boundary
that we have obtained here in the transition between plateau
states.

\acknowledgments We thank A.\ Dobry, E.\ Fradkin, J.\ Polonyi, P.\
Pujol, G.L.\ Rossini and G.\ Silva for helpful discussions. N.E.G.
is grateful to SISSA and ICTP for hospitality and financial support
during the early stages of this work. This work was partially
supported by ECOS-Sud Argentina-France collaboration (Grant A04E03)
and PICS CNRS-CONICET (Grant 18294).


\begin{thebibliography}{99}

\bibitem{von Klitzing} K.\ von Klitzing, G.\ Dorda, M.\ Pepper, Phys.\ Rev.\ Lett.\ {\bf 45}, 494
(1980).

\bibitem{FQHE} D.C.\ Tsui, H.L.\ Stormer, A.C.\ Gossard, Phys.\ Rev.\ Lett.\ {\bf 48}, 1559
(1982).

\bibitem{Laughlin} R.B.\ Laughlin, Phys.\ Rev.\ Lett.\ {\bf 50}, 1395
(1983).

\bibitem{fractional exp} R.\ de-Picciotto, M.\ Reznikov, M.\ Heiblum, V.\ Umansky, G.\
Bunin, D.\ Mahalu, Nature {\bf 389}, 162-164 (1997).

\bibitem{Jain} J.K.\ Jain, Phys.\ Rev.\ Lett.\ {\bf 63}, 199 (1989).

\bibitem{Girvin} S.C.\ Zhang, T.H.\ Hansson, S.\ Kivelson, Phys.\ Rev.\ Lett.\ {\bf 62} (1989) 82; S.\ Girvin
in R.E.\ Prange and S.M.\ Girvin, The Quantum Hall Effect
[Springer-Verlag, 1990]; E.\ Fradkin, Field Theories of Condensed
Matter Systems [Addison-Wesley, 1991].

\bibitem{horda} A.\ M.\ Chang, Rev.\ of Mod.\ Phys.\ {\bf 75}, 1449
(2003).

\bibitem{cort} 
  D.~C.~Cabra and N.~E.~Grandi,
  arXiv:cond-mat/0511674.

\bibitem{Wen} X-G.\ Wen, Adv.\ in Physics, {\bf 44}, 405
(1995).

\bibitem{first diff} M.\ Grayson {\it et al}, Phys.\ Rev.\ Lett.\ {\bf 80},
1062 (1998).

\bibitem{Yang} K. Yang, Phys. Rev. Lett. 91, 036802 (2003).

\bibitem{Wan} X.\ Wan, F.\ Evers, E.H.\ Rezayi, Phys.\ Rev.\ Lett.\
{\bf 94}, 166804 (2005).

\bibitem{plateau} A.\ M.\ Chang {\it et al}, Phys.\ Rev.\ Lett.\ {\bf 86},
143 (2001).

\bibitem{exp-num} A.\ W\"urtz {\it et al}, Phys.\ Rev.\ {\bf B 65}, 075303 (2002);
Y.\ N.\ Joglekar {\it et al}, Phys.\ Rev.\ {\bf B 68}, 035332
(2003); D.\ B.\ Chklovskii , Phys.\ Rev.\ {\bf B 46}, 4026 (1992);
 D.\ B.\ Chklovskii, Phys.\ Rev.\ {\bf B 51}, 9895 (1995).

\bibitem{Chamon-Fradkin} C.\ de C.\ Chamon, E.\ Fradkin, Phys.\ Rev.\ {\bf B
56}, 2012 (1997).

\bibitem{nonuniv} C.\ de C.\ Chamon, X-G.\ Wen, Phys.\ Rev.\ {\bf B 49}, 8227
(1994);  D-H Lee, X-G Wen, arXiv:cond-mat/9809160v2

\bibitem{NR} {Numerical Recipies},  William H Press, Saul A. Teukolsky, William T.
Vetterling, Brian P. Flannery, Cambridge University Press, ISBN 0521750342

\bibitem{Susskind} L.\ Susskind, preprint hep-th/0101029.

\bibitem{CG} D.\ C.\ Cabra, N.\ Grandi, in preparation.

\bibitem{Poly}  A.\ P.\ Polychronakos, JHEP {\bf 0104}, 011 (2001).


\end{thebibliography}
\end{document}